\documentclass[useAMS,usenatbib]{mn2e}
\usepackage{latexsym,amsmath,amssymb,enumerate}
\usepackage[]{graphicx,lscape}
\usepackage{subfigure}
\usepackage{multirow}

\usepackage{color,verbatim}
\usepackage{ulem}

\DeclareMathOperator{\erf}{erf}

\title[Resolving the Mass--Anisotropy Degeneracy]{ 
Resolving the mass--anisotropy degeneracy of the spherically symmetric Jeans equation I: theoretical foundation}
\author[F. I. Diakogiannis, G. F. Lewis and R. A. Ibata]{
Foivos I. Diakogiannis$^{1}$\thanks{E-mail:
f.diakogiannis@physics.usyd.edu.au}, Geraint F. Lewis$^{1}$    and Rodrigo A. Ibata$^{2}$\\
$^{1}$Sydney
Institute for Astronomy, School of Physics, A28, University of Sydney, NSW 2006, Australia\\
$^{2}$Observatoire
Astronomique, Universit\'{e} de Strasbourg, CNRS, 11, rue de l Universit\'{e}, F-67000 Strasbourg, France}
\begin{document}

\pagerange{\pageref{firstpage}--\pageref{lastpage}} \pubyear{2013}

\maketitle

\label{firstpage}
\begin{abstract}
A widely employed method for estimating the mass of stellar systems with apparent spherical symmetry is dynamical modelling using the spherically symmetric Jeans equation. 
Unfortunately this approach suffers from a degeneracy between the assumed mass density and the second order velocity moments. This degeneracy can lead to significantly different predictions for the mass content of the system under investigation, and thus poses a barrier for accurate estimates of the dark matter content  of astrophysical systems.   In  a series of papers  we describe an algorithm that removes this degeneracy and allows for unbiased mass estimates of systems of constant or variable mass-to-light ratio.
The present contribution sets the theoretical foundation of the method that  reconstructs a unique kinematic profile for some assumed free functional form of the mass density. 
The essence of our method lies in 
using  flexible B-spline functions for the representation of the radial velocity dispersion in the spherically symmetric Jeans equation.  
We demonstrate our algorithm through an application to synthetic data for the case of an isotropic King model with fixed mass-to-light ratio, recovering excellent fits of theoretical functions to observables and a unique solution. 
The mass-anisotropy degeneracy is removed to the extent that, for an assumed functional form of the potential and mass density pair $(\Phi,\rho)$, and a given set of line-of-sight velocity dispersion $\sigma_{los}^2$ observables, we recover a unique  profile for $\sigma_{rr}^2$ and $\sigma_{tt}^2$.  
Our algorithm is simple, easy to apply and provides an efficient means to reconstruct the kinematic profile.
\end{abstract}


\section{Introduction}

The spherically symmetric Jeans equation (hereafter SSJE) is an important tool for the estimation of the mass content of stellar structures that exhibit spherical symmetry. It has been used widely \citep[see][]{2008gady.book.....B} for the dynamical modelling of globular clusters, dwarf spheroidal and elliptical galaxies with nearly spherical shape. However, a problem with this approach is that there exists a degeneracy between the assumed mass density  and the velocity distribution of the system, which can lead to erroneous mass estimates. Describing  the mass content of a stellar system accurately is crucial for identifying dark matter (hereafter DM) structures and to test the standard $\Lambda$CDM model. Therefore it would be important if this degeneracy could be completely removed.

There has been extensive effort 
\citep[e.g.][and others]{1982MNRAS.200..361B,1983ApJ...266...58T,
1987ApJ...313..121M,1990AJ.....99.1548M,1992ApJ...391..531D,2002MNRAS.333..697L}
to resolve this problem in recent years with significant, yet not complete, success.  There are two main approaches in attacking the problem. 
One approach is to assume a functional form for the mass density and then try to recover the correct second order velocity moments. 
The other is to define a class of distribution functions  $f(E,L)$ and try to infer qualitative and quantitative results for the velocity distribution of actual stellar systems though the use of second, fourth or higher velocity moments of the observables.   
It should be stated that both approaches try to estimate a unique kinematic profile for a given mass density. Both have advantages and disadvantages. The first has the advantage that we can make a good prediction of the functional form of the mass density from observed brightness distributions.  However this method is limited by the use of only the second velocity moments, thus it
 cannot account for the general velocity distribution. The second approach can, in principle, estimate the full distribution function (hereafter DF). However we do not have a direct comparison of $f(E,L)$ with $E,L$ observables to be certain of our assumption on the functional form of $f(E,L)$. Thus it is possible that it introduces a bias in the derived measures.

In the present paper we focus on the first approach. The seminal work of  \citet[][see also \citealt{1983ApJ...266...58T}]{1982MNRAS.200..361B}  presented an algorithm that,  for self consistent systems and an assumed mass density, yields a unique constant mass-to-light ratio   $\Upsilon$  and second radial $\sigma_{rr}^2$ and tangential $\sigma_{tt}^2$ velocity moments.  Although this method is elegant, significant and very useful, it presents some difficulties and has limitations. The major  limitation, as the authors point out,  is that it cannot account for a variable mass-to-light ratio; i.e.  when a separate dark matter component is present the method is not applicable. Another difficulty  is that the accuracy of the  algorithm was demonstrated using synthetic data that had very small errors ($\leq 3 \%$ of the actual values), which is rarely realistic in practise.
Furthermore,  one needs to define first a fitted profile to the line-of-sight velocity dispersion $\sigma_{los}^2$ and then use this as a theoretical function to infer the functional form of the anisotropy $\beta$. This $\sigma_{los}^2$ fit will always have some uncertainty due to errors in the observables.
The authors demonstrate that this uncertainty  does not affect their qualitative results, i.e. they can still distinguish between radially or tangentially biased profiles. Unfortunately quantitatively, with such a procedure, there is error propagation that degrades the quality of the  estimates, particularly if there are large uncertainties in the data. It was argued by \cite{1994MNRAS.270..271V} that this method requires knowledge of the profile of the  projected velocity dispersions corrected for the effects of seeing and spatial binning and that these corrections can be exceedingly difficult to make, especially near the centre of the system under study.

\cite{2013MNRAS.428.3648I}  presented an algorithm for the evaluation of the mass content of a system with variable mass-to-light ratio and a varying anisotropy profile; i.e. the method can account for a separate dark matter component.  This method uses splines to define the radial $\Delta\sigma_{rr}^2(r_i)$ and tangential $\Delta\sigma_{tt}^2(r_i)$ velocity dispersions,  as well as the mass density  $\Delta\rho(r_i)$ in a dense set of radial positions $r_i$. The individual value of each profile at each position $r_i$ is treated as a free parameter and  is estimated through an MCMC scheme subject to some physically plausible constraints. This method, although efficient,  uses a large number of free parameters ($\sim 389$ free parameters) and is computationally expensive, thus making model comparison through Bayesian model inference a very difficult task.

Our work focuses on the task of determining unique second order velocity moments  and accurate mass estimates performed using the SSJE. 
In the present paper we  develop  the basic mathematical framework of our algorithm. Thus we limit the application of our method to a simple example of a system with a fixed mass-to-light ratio. In  (submitted to MNRAS MN-14-0102-MJ; hereafter Paper II) we expand the theoretical model and validate our method by giving a detailed analysis of applications to various systems with constant and variable mass-to-light ratio.  In the current approach, 
the only assumption we make is the functional form of the mass density $\rho(r)$ of the system. From this, facilitating comparison with observables, we recover the mass content and  a unique kinematic profile of the stellar system. Then the correct mass model hypothesis can be inferred through Bayesian inference methods. 
Our method is valid even in the case where there are two separate components, e.g.. stars and DM (Paper II). It is simple, easy to apply and computationally inexpensive. The key idea behind our method is this: 
the line-of-sight velocity dispersion, $\sigma_{los}^2$,  depends on both the radial, $\sigma_{rr}^2(r)$, and tangential,  $\sigma_{tt}^2(r)$, velocity dispersions; 
since we do not know the functional form of the kinematic quantities $\sigma_{rr}^2(r)$ or $\sigma_{tt}^2(r)$,  we can use the SSJE to eliminate the tangential component, $\sigma_{tt}^2(r)$, dependence from $\sigma_{los}^2$ and 
approximate  $\sigma_{rr}^2(r)$ with a smooth Computer Aided Geometric Design (CAGD) curve.   
  Comparison of   $\sigma_{los}^2$ with line-of-sight velocity dispersion observables  gives $\sigma_{rr}^2(r)$  both the correct geometric shape and estimates of its numerical value. This avoids any bias in the mass estimates from assumption of a specific anisotropy profile. The CAGD tools we use are B-spline functions. Once $\sigma_{rr}^2(r)$ is known, we  can always use the SSJE to estimate the tangential velocity dispersion, $\sigma_{tt}^2(r)$, thus recover, within uncertainties, the anisotropy profile.

The structure of our paper is the following: in section \ref{DLI_BSplines_Jeans_detail} we describe the degeneracy of the SSJE in a detailed mathematical formulation. 
In section \ref{DLI_BSplines_MathForm} we give an extended presentation of smoothing B-spline CAGD curves and functions, how we combine them with the SSJE and the dynamical mass model we use. In section \ref{DLI_BSplines_StatAnal} we describe the statistical inference methods. In section \ref{DLI_BSplines_Examples} we present a simple  example. In this we reconstruct fully the mass content of the system and the kinematic profile, using synthetic  data of brightness and line-of-sight velocity dispersion $\sigma_{los}^2$.  In section \ref{DLI_BSplines_Discussion}
we discuss various aspects of our method, and we comment on the optimum smoothing problem of the B-spline representation. Finally in section \ref{DLI_BSplines_Conclusions}
 we conclude our work.

\section{Jeans degeneracy in detail}\label{DLI_BSplines_Jeans_detail}

Consider a self gravitating stellar system in dynamical equilibrium. Under the SSJE framework this system is described through the mass density $\rho(r)$, the potential\footnote{For self consistent systems, potential $\Phi(r)$ and mass density $\rho(r)$ are related through  Poisson's equation.} $\Phi(r)$ and the second velocity moments $\sigma_{rr}^2(r)$ and $\sigma_{tt}^2(r)$. 
The SSJE is customarily written in the form: 
\begin{equation} \label{DLI_BSplines_Jeans_traditional}
-\frac{d\Phi}{dr} = \frac{1}{\rho} \frac{d (\rho \sigma_{rr}^2)}{dr} 
+ \frac{2 \beta}{r} \sigma_{rr}^2
\end{equation}
where  
\begin{equation}\label{DLI_Binney_beta}
\beta \equiv 1 -\frac{\sigma_{tt}^2}{2 \sigma_{rr}^2} 
\end{equation}
is the Binney  anisotropy parameter\footnote{Here we consider that in a spherical coordinate system $(r,\theta,\phi)$, the tangential velocity dispersion is defined as $\sigma_{tt}^2=\sigma_{\theta\theta}^2+\sigma_{\phi\phi}^2$.} \cite[][ see also \citealt{2008gady.book.....B}]{1982MNRAS.200..361B}. The connection with observables is performed through the line-of-sight velocity dispersion, namely: 
\begin{equation}
\label{DLI_BSplines_sigmalos_traditional}
\sigma_{los}^2(R) = \frac{2}{\Sigma(R)} \int_{R}^{r_t} 
\left( 1-\beta(r) \frac{R^2}{r^2} \right) \frac{r \; \rho \; \sigma_{rr}^2}{\sqrt{r^2-R^2}} dr
\end{equation}
where $r_t$ is the tidal radius of the physical system. Note that $\beta$  is multiplied with $\sigma_{rr}^2$, and this increases the complexity of the set of Equations  \ref{DLI_BSplines_Jeans_traditional} and \ref{DLI_BSplines_sigmalos_traditional}.

The traditional approach of using the SSJE for dynamical modelling is to assume a mass density $\rho(r)$ and a functional form for  the $\beta(r)$ anisotropy profile. Then one evaluates $\sigma_{rr}^2(r)$ from Equation \ref{DLI_BSplines_Jeans_traditional}, substitutes into Equation \ref{DLI_BSplines_sigmalos_traditional} and compares with observables.  For an assumed mass density any 
 $\beta(r)$ functional form defines a severe restriction on the system and inserts bias in the mass estimates. Choosing different $\beta(r)$ functions in general can result in significantly different results  for both the mass estimates and the kinematic profile of the system \citep{1987ApJ...313..121M}.

 As mentioned in the introduction this is the problem we are going to resolve: for an assumed mass density $\rho(r)$ we will recover the unique kinematic profile as it is described through the second moments of radial $\sigma_{rr}^2$ and tangential $\sigma_{tt}^2$ velocities. We must emphasize that this does not remove the degeneracy on the assumption of the mass density, i.e. 
 a different assumption on $\rho(r)$  will in general lead to a different kinematic profile $\sigma_{rr}^2(r)$ and $\sigma_{tt}^2(r)$ that still reproduces the observables.  However, again this will be unique for the given $\rho(r)$.

\section{Mathematical  formulation }\label{DLI_BSplines_MathForm}
Since B-spline functions are not widely used in the astronomical community, we will give a short description of them. In this section we will introduce B-spline curves and functions and  describe in detail how we use B-spline functions in the spherically symmetric Jeans equations. We will also  give definitions for the mass density $\rho(r)$ of  the dynamical models we use. 
The standard reference for B-spline functions is \cite{de1978practical}. 
For practical applications the interested reader will find great help in books of Computer Aided Geometric Design (CAGD),  such as \cite{Rogers:2001:INH:347021} and \cite{Farin:2001:CSC:501891}\footnote{There are also some excellent online notes by C. K. Shene  http://www.cs.mtu.edu/$\sim$shene/COURSES/cs3621/NOTES/}. All the above references provide information on available libraries for B-splines in FORTRAN and C programming languages. For our needs we used the GNU Scientific Library (GSL) that has an implementation of B-spline bases. 

In  short, a B-spline function $f(x)$ is a linear combination of some constant coefficients $a_i$ with some polynomial functions $B_{i,k}(x)$  (B-spline basis functions) of a given degree ($k-1$), i.e. $f(x) = \sum_i a_i B_{i,k}(x)$. These polynomial functions $B_{i,k}(x)$ are smooth and consist of polynomial pieces joined together in a special way. We will start with the definition of B-spline basis functions and then proceed to B-spline curves and functions.

\subsection{B-spline basis}
Let $k$ be a positive integer and $\xi_i$ represent a non-decreasing sequence of $m+1$ real numbers, $\xi_0 \leq \xi_1 \leq \cdots \leq \xi_m$.  We will refer to this sequence as the {\bf knot} sequence. Each of these $\xi_i$ are called knots. The integer $k$ is called the {\bf order} of the B-spline basis and should not be confused with the degree of the polynomial pieces (degree  $=k-1$).  
We say that a knot $\xi_i$ has multiplicity $p$ if it appears $p$ times in the knot sequence 
($p\leq k$).

The elements $B_{i,1}(x)$ of a   B-spline basis of order 1 (polynomial degree $=0$) are defined through the formula: 
\begin{equation}
B_{i,1}(x) =
\begin{cases}
 1, \quad \text{if}\quad \xi_i \leq x < \xi_{i+1}\\
 0, \quad \text{otherwise}
\end{cases}
\end{equation} 
 A B-spline basis  of order $k$ is defined for all real numbers $x$ through the Cox--de Boor  recursive algorithm: 
\begin{equation}
\label{DLI_BSplines_definition1}
B_{i,k}(x) = \omega_{i,k} B_{i,k-1}(x) 
+ \left(1-\omega_{i+1,k} \right) B_{i+1,k-1}(x)
\end{equation}
 where 
 \begin{equation}
 \omega_{i,k}(x) = 
 \begin{cases}
\frac{x-\xi_i}{\xi_{i+k-1} - \xi_i}, \quad \text{if} \quad 
 \xi_{i+k-1} \neq \xi_i\\
     0, \quad \text{otherwise}
 \end{cases}
 \end{equation}
 Thus  B-spline basis functions $B_{i,k}(x)$ are polynomials of degree $k-1$. 
 In this definition we follow the convention that whenever division 
 by zero appears  we treat the whole fraction as zero, i.e. $0/0 \equiv 0$.  

 We list here some important properties of the B-spline basis  which are related to our needs for the development of our method. This is not a complete list. In our effort to emphasize the importance of these properties in applications, we shall frequently refer to the coefficients $a_i$ of a B-spline function  $f(x)=\sum_i a_i B_{i,k}(x)$,  despite the fact that we formally define these functions in a later subsection:   
 \begin{enumerate}
 \item B-spline basis functions $B_{i,k}(x)$ are linearly independent. 
 \item $B_{i,k}(x)$ is a degree $k-1$ polynomial in $x$. This is a restriction on the differentiability of the functions we are going to consider later.  
 \item Each basis function $B_{i,k}(x) \geq 0$ for any $x$. Then, any change in sign of a B-spline function results from a change in the sign of the coefficients $a_i$. This is a very important property, since if we have a positive function (such as $\sigma_{rr}^2$) that we wish to expand in a B-spline basis, then by demanding the coefficients $a_i$ of this expansion to be positive, we guarantee this restriction. 
 \item For a given knot sequence $\xi_0,\ldots,\xi_m$ there exist $n$ B-spline basis  functions $B_{i,k}(x)$, of order $k$, where $n=m+1-k$. If we wish to use a given polynomial order B-spline basis, 
 and a given number of coefficients $a_i$, the number of knot points is uniquely determined.  
 \item On any point $x \in [ \xi_i, \xi_{i+1})$ at most $k$ basis functions are non zero.
 Then for a B-spline function $f(x)=\sum_i a_i B_{i,k}(x)$ for a given $x \in  [\xi_i,\xi_{i+k})$, only a subset of all coefficients $a_i$ will contribute to the value of $f(x)$. We shall refer to this property as the local modification scheme of B-splines.  
  \end{enumerate}

\begin{figure}
\centering
\includegraphics[width=\columnwidth]{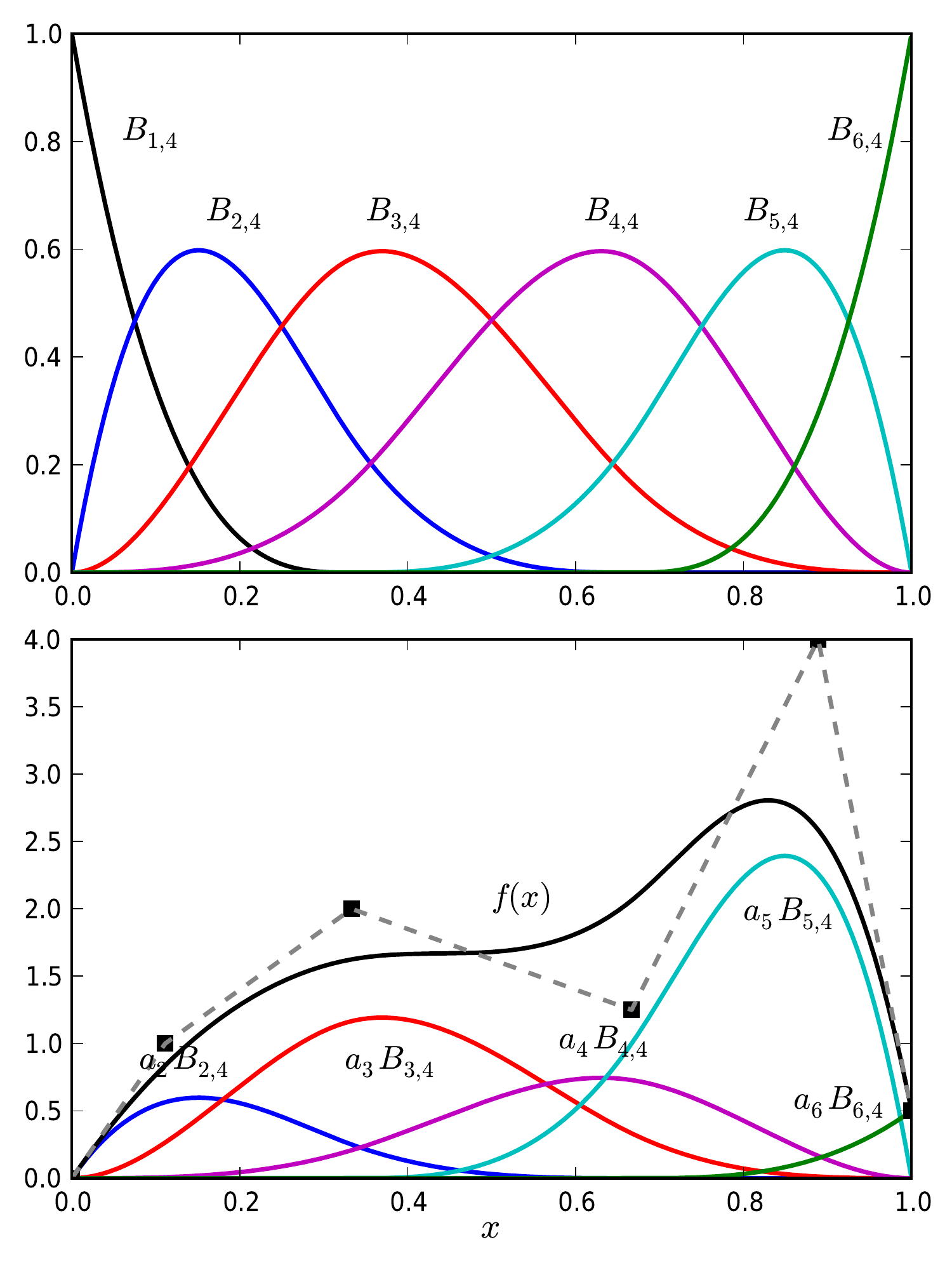}
 \caption{Top panel: B-spline basis functions of order $k=4$ (polynomial degree $=3$) for the uniform knot sequence $[0,0,0,0,1/3,2/3,1,1,1,1]$. Bottom panel: a B-spline representation of a function $f(x)= \sum_i a_i B_{i,k}(x)$ for the set of coefficients  $a_i = \{0,1,2,1.25,4.0,0.5 \}$ on the same knot sequence. The B-spline basis functions multiplied with the corresponding coefficient are also plotted. The control polygon is the dashed line, and the black squares are the positions of the control points. Due to the multiplicity of the first and last knots the function $f(x)$ attains the values of the first $a_1=0$ and last $a_6=0.5$ coefficients. }
\label{DLI_BSplines_Fig1}
\end{figure} 
In Fig. \ref{DLI_BSplines_Fig1}  we plot (top panel) the B-spline basis functions $B_{i,4}(x)$ (order $k=4$, polynomial of degree $k-1=3$)  for the knot sequence $\xi_i=[0,0,0,0,1/3,2/3,1,1,1,1]$. The dimension of the knot vector is  $\dim (\bxi)=10$, thus according to our definition $m=9$. Then there exist $n = m+1-k = 6$ linearly independent bases of order $k=4$. 

\subsection{B-spline Curves} 
A B-spline curve in 2 dimensional space is the linear combination of some $n$ constant 
 vector coefficients\footnote{Points in 2D space for our needs.} $\mathbf{c}_i$  with the B-spline basis functions $B_{i,k}$: 
\begin{equation}
\mathbf{P}(x) = \sum_{i=1}^{n} \mathbf{c}_i B_{i,k}(x)
\end{equation} 
$\mathbf{P}(x)$ is the position vector that traces the curve parametrized by  $x$. The position vectors $\mathbf{c}_i$ are called {\bf control points}. They define the $n$ vertices of  an  open polygon which is called the {\bf control polygon}. This control polygon defines the shape of the B-spline curve. By adjusting the control points, the curve acquires a different geometric shape.  

We state without proof two very important properties of B-spline curves: 
\begin{enumerate}
\item An important class of B-spline curves is the one for which the first $\xi_0$ and last $\xi_m$ knots have multiplicity $p$ equal to the order $k$ of the B-spline curve. It can be proved then, that the B-spline curve passes from the first $\left(\mathbf{c}_1=\mathbf{P}(\xi_0)\right)$ and last $\left(\mathbf{c}_n=\mathbf{P}(\xi_m)\right)$ points of the control polygon. This is crucial for our subsequent analysis since, if from some physical considerations we know the boundary conditions at the beginning or the end of a curve, then 
we know the coordinates of the first $\mathbf{c}_1$ or last $\mathbf{c}_m$ control points. 
This property in combination with the smooth behaviour of B-spline curves proves to be a  severe restriction on our models. 
 All curves we are going to consider have multiplicity $p=k$ in the first and last knots.       
\item Having defined a knot vector $\bxi$ and knowing the control points $\mathbf{c}_i$, then these completely determine the tangent curve   curve $\mathbf{T}(x)$ of $\mathbf{P}(x)$. This is simply:
\begin{equation}
\mathbf{T}(x) = \frac{d \mathbf{P} (x)}{dx} = \sum_{i=1}^n \mathbf{c}_i \frac{d B_{i,k}(x)}{dx}
\end{equation}
Since the basis functions $B_{i,k}(x)$ are known polynomial functions, so are their derivatives. Thus,  the control points $\mathbf{c}_i$ define the curve and all of its derivatives. This is a remarkable property for our needs in dynamical analysis. Each time we encounter an unknown function that participates in some differential equation, then by using a B-spline representation of the function we no longer need to solve the differential equation. Instead, we simply need to calculate the unknown coefficients $a_i$ through some algebraic process\footnote{This applies to differential equations of the form: 
\[
\lambda_n(x) \frac{d^n y(x)}{dx^n} + \cdots + \lambda_1(x) \frac{d y(x)}{dx}
+\lambda_0(x)=0,
\]
where $\lambda_i(x)$, $i=0,\ldots,n$ are arbitrary functions of $x$, but not $y$.
}.  We shall see later that this property removes the complexity in the SSJE of having to calculate $\sigma_{rr}^2$ and its first derivative. 
\end{enumerate}

\subsection{B-spline Functions}
\label{DLI_paperI_BSpline_functions}
A B-spline function is the linear combination of some constant coefficients $a_i$ with the B-spline basis functions $B_{i,k}(x)$:   
\begin{equation}
f(x) = \sum_{i=1}^{n} a_i B_{i,k}(x)
\end{equation}  
The properties of B-spline curves are transfered also to B-spline functions: 
\begin{enumerate}
\item For a B-spline function $f(x) = \sum_{i=1}^n a_i B_{i,k}(x)$ defined on some knot vector, if the multiplicity of the first $\xi_0$ and last $\xi_m$ knot is equal to the B-spline basis order $k$ then $f(\xi_0) = a_1$ and $f(\xi_m) = a_n$. 
\item For a given knot sequence $\xi_0,\ldots,\xi_m$, the constant coefficients $a_i$ uniquely determine the function $f(x)$ and all of its derivatives. 
\end{enumerate}

An example of a B-spline function is given  in the bottom panel of Fig. \ref{DLI_BSplines_Fig1}.   We define our function with the use of the   
 B-spline basis functions $B_{i,4}(x)$ that are on the top panel (order $k=4$,   knot sequence $\xi_i=[0,0,0,0,1/3,2/3,1,1,1,1]$)
and the set of coefficients $a_i = \{0,1,2,1.25,4.0,0.5 \}$; we plot  the function $f(x) = \sum_{i=1}^{6} a_i B_{i,4}(x)$, the weighted B-spline basis functions $a_i B_{i,4}(x)$ as well as the control polygon of the B-spline curve $\mathbf{P}(x) = (x,f(x))$. The coordinates of the control points are given by $\mathbf{c}_i = \binom{a_i}{\xi_i^{*}}$, where $\xi_i^{*}$ are called Greville abscissae and are  not to be confused with the knot points $\xi_i$.
These are defined as the mean position of $k-1$ consecutive knots $\xi_i$
\[
\xi_i^{*} = \frac{1}{k-1} (\xi_{i}+\xi_{i+1}+\cdots + \xi_{i+k-2})
\]
see \cite{Farin:2001:CSC:501891} for details.

B-spline curves and functions are used extensively in CAGD and in statistical modelling of data, whenever a smoothing model function is needed. The quality of the resulting fit depends on the  order $k$ of the spline, on the distribution of knot points\footnote{In general we want more control points around regions of $x$ where the function we wish to model has greater curvature.}, and on the number of coefficients. There is no optimum choice since all of the above parameters depend on our data. We need to use model comparison  for the best choices of order $k$, knot distribution and number of knot points. In general, a bad choice of all the above parameters can result in overfitting or underfitting to the data. Bayesian inference solves partially this problem by finding the model that has the optimum knot order $k$ and number of coefficients $a_i$.  Again, there still remains the problem of optimum smoothing, since it may be the case that we have data with large errors that result in unphysical oscillatory behaviour in the functions we represent with B-spline bases. We give a solution to this in Paper II  by introducing a smoothing penalty that uses information of the smoothness from ideal theoretical models. 

 Our goal is to use a B-spline function representation for the radial velocity dispersion $\sigma_{rr}^2(r)$: 
\[
\sigma_{rr}^2(r) = \sum_{i=1}^n a_i B_{i,k}(x).
\] 
Doing so, we recover the values of the coefficients $a_i$ from comparison with observational values of the line-of-sight velocity dispersion $\sigma_{los}^2(R)$, thus determining the anisotropy of the system in a unique way.

\subsection{Choice of knot sequence} 
\label{DLI_BSplines_knot_section}
\begin{figure}
\centering
\includegraphics[width=\columnwidth]{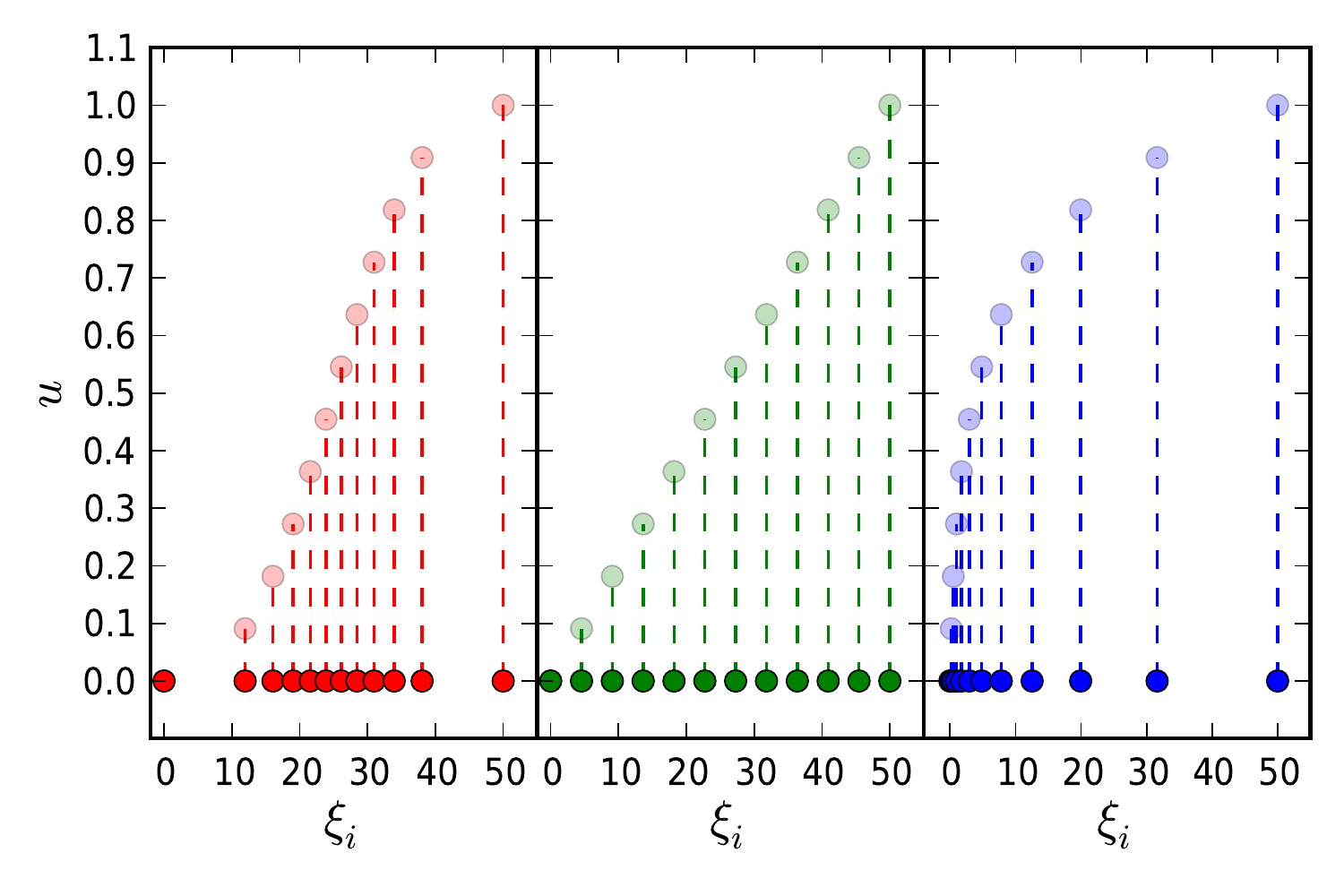}
 \caption{ Various knot distributions for a B-spline basis. 
 Left panel: Gaussian knot distribution $\xi_i$ around mean $\mu = 25$ with $\gamma=5$. Middle panel: uniform knot distribution. Right panel: exponential knot distribution.}
\label{DLI_BSplines_Fig_knots}
\end{figure} 

The distribution of knot points $\xi_i$ is one way to affect the geometric shape of a curve described by a B-spline function\footnote{Others are the choice of order $k$ and the coefficients $a_i$.} $\mathbf{P}(x) = (x,f(x))$. For our purposes we want to approximate a physical quantity, i.e. $\sigma_{rr}^2(r)$, with a B-spline representation. This approximation is better if we have more knot points distributed around regions where our function has greater curvature. If we have no information on where this region might be, we use a uniform distribution of knot points.

For a King mass model,  we know that a critical distance from the cluster center is the King core radius $r_c$. It is around this point where the several functions of the model appear to have increased curvature. Therefore,  we have the option of using a Gaussian knot distribution with mean $\mu = r_c$ and variance $\sigma = r_t  / \gamma$,  
where $r_t$ is the tidal radius of our system. 
The parameter $\gamma$ regulates how close to the mean the distribution of points will be. A large value of $\gamma$ concentrates points around $r_c$. A value $\gamma = 1$ results in an approximately uniform distribution in the interval $\xi_i \in [0,r_t]$.

Let  $u$  be a  uniform  sequence of numbers in the interval $[0,1]$. For this sequence, the following equation gives rise to a Gaussian distribution of points around the mean $\mu$ with variance $\sigma$:
\begin{multline} \label{DLI_BSplines_gauss_knot}
\xi(u) = \mu + \sqrt{2} \sigma \erf^{-1} \biggl[
\erf \left(\frac{ \mu}{\sqrt{2} \sigma}\right) (u-1)+
\\ 
+u 
\erf \left(\frac{r_t - \mu}{\sqrt{2} \sigma}\right) 
\biggr]
\end{multline}
Equation \ref{DLI_BSplines_gauss_knot} is produced with the same methodology we use when we wish to create a random Gaussian number $\xi \in [0,r_t]$ from a uniform random $u \in [0,1]$.  

In Fig. \ref{DLI_BSplines_Fig_knots} we plot several possible choices of knot distributions. Specifically the left panel demonstrates a Gaussian distribution of points around mean $\mu = 25$ with coefficient $\gamma=5$. The middle panel has a uniform distribution of knots, while the right panel is an exponential knot distribution with the majority of knots concentrated exponentially close to the origin.

\subsection{The Spherically Symmetric Jeans  Equation} \label{DLI_BSplines_Jeans_section}
In this section we  describe how we combine the SSJE with the line-of-sight velocity dispersion $\sigma_{los}^2$ in order to facilitate comparison with observables. For the case of our Galaxy, where  typically, one only has radial velocities, see Appendix \ref{DLI_BSplines_MW}.

In order to apply our method, we write  Equations \ref{DLI_BSplines_Jeans_traditional} and \ref{DLI_BSplines_sigmalos_traditional} in what we believe to be a much simpler form in terms of $\sigma_{rr}^2$ and $\sigma_{tt}^2$. Furthermore, we simplify the notation  by setting: 
\begin{align}
\psi &=  \sigma_{rr}^2(r)\\
\phi &=  \sigma_{tt}^2(r).  
\end{align}
Then  the SSJE  and $\sigma_{los}^2(R)$ in the $\psi$, $\phi$ representation are:
\begin{align}\label{DLI_BSplines_Jeans_psi}
- \frac{d \Phi}{dr} & = \frac{d\psi}{dr} 
+ \left( \frac{1}{\rho} \frac{d \rho}{dr} +\frac{2}{r} \right) \psi - \frac{1}{r} \phi\\ 
\sigma_{los}^2 & = \frac{1}{\Sigma(R)} \label{DLI_BSplines_sigmalos_psi}
\int_{R}^{r_t} \frac{\rho \left[ 2 \psi (r^2-R^2) + \phi R^2 \right]}{r \sqrt{r^2-R^2}}
\end{align}  
As we shall see in section \ref{DLI_BSplines_DynModel} the tidal radius $r_t$ of the system is defined through Poisson's equation from $\rho(r)$ and $\Phi(r)$ and does not depend on the kinematic quantities $\sigma_{rr}^2$ or $\sigma_{tt}^2$.
The problem with 
Equations \ref{DLI_BSplines_Jeans_psi} and \ref{DLI_BSplines_sigmalos_psi} is that both functions $\psi$ and $\phi$ are unknown, and cannot be deduced from the mass density $\rho(r)$ or the potential $\Phi(r)$ of the system. Moreover, $\psi$ participates also with its first derivative, making the problem even more complex. 

We are going to consider the expansion of $\psi(r)$ in a B-spline basis function of order $k$. That is: 
\begin{equation}\label{DLI_BSplines_BasisExpansion}
\psi(r) = \sum_{i=1}^{N_{\text{coeffs}}} a_i B_{i,k}(r)
\end{equation}
where $B_{i,k}(r)$ are known B-spline basis functions. 
Then the derivative of this function is merely: 
\begin{equation} \label{DLI_BSplines_dpsi}
  \psi^{(1)}(r) =  \sum_{i=1}^{N_{\text{coeffs}}} a_i  B^{(1)}_{i,k}(r)
\end{equation}
where $\psi^{(1)}(r) = d \psi/dr$ and $B^{(1)}_{i,k}(r) = d B_{i,k}(r)/dr$. 
That is, the derivative of $\psi$ depends on the same unknown coefficients $a_i$ but is expanded in a  new set of basis  functions $B_{i,k}^{(1)}(r)$. This removes the complexity of not knowing the derivative of $\psi(r)$. 
Substituting $\phi$ from Equation \ref{DLI_BSplines_Jeans_psi} in  the integrand of $\sigma_{los}^2$ (Equation \ref{DLI_BSplines_sigmalos_psi}) yields: 
\begin{multline}
\sigma_{los}^2  = \frac{1}{\Sigma(R)}\int_R^{r_t}\frac{\left(2 r \rho + \rho^{(1) }R^2\right)  \psi + \rho  R^2 \psi^{(1)}  }{\sqrt{r^2-R^2}} dr \\ 
+
\frac{1}{\Sigma(R)}\int_R^{r_t}
 \frac{\rho R^2}{\sqrt{r^2-R^2}} \frac{d \Phi}{dr} dr
\end{multline}
where  $\rho^{(1)} = d\rho(r)/dr$. 
Now the line-of-sight velocity dispersion depends on the mass density of the system, the potential and the unknown function $\psi$ along with its first derivative $\psi^{(1)}$. Using the basis expansion (Equations \ref{DLI_BSplines_BasisExpansion} and \ref{DLI_BSplines_dpsi}) yields:
\begin{multline}
\sigma_{los}^2=\\ 
\sum_i a_i \frac{1}{\Sigma(R)} \int_R^{r_t}
\frac{\left(2 r \rho + \rho^{(1) }R^2\right)  B_{i,k}(r) + \rho  R^2 B_{i,k}^{(1)}(r)  }{\sqrt{r^2-R^2}}dr \\
+\frac{1}{\Sigma(R)} \int_R^{r_t} \frac{\rho R^2}{\sqrt{r^2-R^2}} \frac{d \Phi}{dr} dr
\end{multline}
We define the following functions: 
\begin{align} \label{DLI_BSplines_I}
I_{i} (R) &\equiv \frac{1}{\Sigma(R)}
\int_{R}^{r_t}\frac{\left(2 r \rho + \rho^{(1) }R^2\right)  B_{i,k}(r) 
+ \rho  R^2 B_{i,k}^{(1)}(r)  }{\sqrt{r^2-R^2}}\\
\label{DLI_BSplines_C}
C(R) &\equiv\frac{1}{\Sigma(R)}\int_{R}^{r_t} \frac{\rho(r) R^2}{\sqrt{r^2-R^2}} \frac{d \Phi(r)}{dr}
\end{align}
Then,  the value of $\sigma_{los}^2 (R)$ is given by:
\begin{equation} \label{DLI_BSplines_fundamental}
\sigma_{los}^2(R)  = \sum_i a_i I_i(R) + C(R)
\end{equation}  
Comparing  the line-of-sight velocity dispersion $\sigma_{los}^2$ 
with observables we can determine the marginalized distributions of the unknown coefficients $a_i$ as well as the defining parameters of the mass model. That is, although we cannot  know the velocity profile of the cluster from its mass density $\rho(r)$ or the potential, we may allow this to be deduced from the observables. Knowledge of $a_i$ is equivalent to knowledge of $\sigma_{	rr}^2\equiv\psi$ and $\sigma_{tt}^2\equiv \phi$.

The coefficients $a_i$ cannot take arbitrary values. One restriction to be applied is that both $\psi\equiv\sigma_{rr}^2$ and $\phi$ functions (Eq. \ref{DLI_BSplines_Jeans_psi}) are positive. Moreover, the function $\psi$, must equal zero at $r=r_t$, the tidal radius of the system,
 and we assume that $\phi$ and $\sigma_{los}^2$ are also zero at $r=r_t$.
This last condition, combined with the smoothness of B-spline functions, imposes a severe restriction on the possible values of $a_i$. The result is well-defined curves with small error bars. We will see later that closer to $r_t$ the variance of the $a_i$ coefficients becomes small. 
 
From Equations  \ref{DLI_BSplines_I} and \ref{DLI_BSplines_C}, we see that $C(r_t)=I_i(r_t)=0$ by definition, since the lower and upper limits of the integrals coincide. Then we must impose an ad hoc restriction that $\sigma_{los}^2 \to 0$ as $r \to r_t$. This is easily achieved by adding an artificial data point very close to $r_t$ in which we demand $\sigma_{los}^2 \approx 0$ within some very small error.

\subsection{Dynamical Models}\label{DLI_BSplines_DynModel}

In the following sections we will reconstruct from synthetic data the kinematic profile of a stellar system in equilibrium, i.e. $\sigma_{rr}^2$ and $\sigma_{los}^2$ (once $\sigma_{rr}^2$ is known, $\sigma_{tt}^2$ can be found from the SSJE). We will assume that the stellar mass content of this system is described by a King-model mass density $\rho(r)$. In the current contribution this is the only mass density  we are going to consider. For systems that contain also a dark matter component, see  Paper II.

For a full description of King models the reader should consult \cite{1966AJ.....71...64K} and 
\cite{2008gady.book.....B}. Here we give for reference the functional forms we used. 
A King model is defined through its distribution function: 
\begin{align}\label{DLI_BSplines_King}
f(\mathcal{E})&=
\begin{cases}
\frac{f_{0} }{ (2\pi \sigma^2)^{3/2} } 
\left(
e^{-\mathcal{E}/\sigma^2} -1
\right) & \mathcal{E}<0 \\
0 & \mathcal{E} \geq 0
\end{cases}
\end{align}
where $f_{0}$ and $\sigma$,  are parameters to be determined from Bayesian  likelihood methods. 

Let $r_t$ denote the tidal radius of the system, i.e. a position beyond  which the mass density and all physical quantities of the system vanish. If $\Phi(r)$ is the potential, by making use of an arbitrary additive constant to its definition, we may define as a new potential the difference: $\Psi=\Phi(r)-\Phi (r_t)$; now $\Psi$ vanishes 
at the tidal radius. Furthermore, in order to simplify our calculations, we introduce the transformation:  
$w=-\Psi(r)/ \sigma^2$.
Then: 
\begin{equation}
\mathcal{E}=\frac{v_r^2+v_t^2}{2}-2\sigma^2 w(r)
\end{equation}
The mass density of the system $\rho(r,w)$ can be calculated analytically with the use of  Computer Algebra Systems (e.g. Maxima, Mathematica, Maple), as functions of radius $r$ and ``potential'' $w(r)$ :
\begin{equation*} 
\rho(r,w)=4 \pi \int_{v_r=0}^{\sqrt{2\sigma^2 w}} 
\int_{v_t=0}^{\sqrt{2\sigma^2 w-v_r^2}} f(\mathcal{E},L)  v_t dv_t dv_r 
\end{equation*}
$v_r$ is the radial component of the velocity in spherical coordinates $(v_r,v_{\theta},v_{\phi})$ and $v_t^2=v_{\theta}^2+v_{\phi}^2$. 
A  model is fully described once we assign values to its defining parameters  and know the functional form of the ``potential'' $w(r)$. 
The latter is achieved by solving Poisson's equation numerically. To do this, we require two additional assumptions at $r=0$: an initial value for the potential $w_0$ and the equilibrium condition $\frac{d w }{dr} \bigr|_{r=0}=0$.

Instead of $(f_{0},\sigma)$ 
 it is very convenient to use  the mass core density $\rho_0$ and the King core radius
$r_c$ defined by:
\begin{align*}
\rho_0& = \rho\left(r,w(r)\right) \bigr|_{r=0}, & 
r_{c} &= \left(
\frac{9 \sigma^2}{4 \pi G \rho_0}
\right)^{1/2}
\end{align*}
Then for the full description of a King  model  we use the following set of parameters $(w_0,\rho_0,r_c)$. 
Using the transformed potential $w$, the Poisson equation is most conveniently written:
\begin{align} \label{DLI_BSplines_Poisson}
 \nabla^2 w(r)&= - \frac{9}{ r_c^2} \tilde{\rho}(r,w), &
 \text{where}\quad 
 \tilde{\rho}(r,w) & = \frac{\rho(r,w)}{\rho_0}
\end{align}
The steps followed for a full evaluation of a King model are the following: 
\begin{enumerate}
\item Assign initial values to parameters $(w_0,\rho_0,r_c)$. 
\item Subject to the initial conditions $w(r=0)=w_0$ and  $\frac{dw}{dr}|_{r=0}=0$,
solve Poisson's equation numerically, to
thus obtain $w(r)$.
\item The mass density $\rho(r)$ is fully determined upon knowledge of $w(r)$. \end{enumerate}

In the following we are going to use only the King mass density, and pretend that we do not know the kinematic quantities $\sigma_{rr}^2$ and $\sigma_{tt}^2$ as defined from the distribution function (Eq. \ref{DLI_BSplines_King}).

\section{Statistical Analysis} \label{DLI_BSplines_StatAnal}
In this section we will be using standard  Bayesian approaches to model fitting.  The reader is directed to standard texts such as \cite{hastie01statisticallearning, sivia2006data} and \cite{2010blda.book.....G}  for further details. 

\subsection{Likelihood function}

Let $\theta$  represent the vector of parameters needed to fully describe a given assumed physical model. These will be the set of defining parameters of the dynamical model, and the coefficients $a_i$ of the B-spline representation of $\psi\equiv\sigma_{rr}^2$, i.e. 
$\theta = (w_0,\rho_0,r_c,a_1,\ldots,a_{n-1})$\footnote{We do not include the last coefficient $a_n$ of the B-spline representation of $\psi=\sigma_{rr}^2$, since this represents the value of $\sigma_{rr}^2(r)$ at the tidal radius $r_t$. As mentioned earlier in the text, the B-spline function $\psi$ passes through the first and last coefficients $a_i$, then $a_n=0$, since $a_n=\sigma_{rr}^2(r_t)=0$.}. 
In the present paper we consider for simplicity an example with fixed mass-to-light ratio $\Upsilon=1$, therefore we do not include $\Upsilon$ as a free parameter. We emphasize however that our algorithm can treat also cases with mass-to-light ratio as a free parameter (Paper II).  
In the framework of Bayesian interpretation we are interested in the posterior probability distribution of these parameters.  Our data set consists of kinematic $D_K$ and brightness $D_B$ data. 
The kinematic data set, $D_K$, consists of line-of-sight velocity dispersion, $\sigma_{los}^2$, values that can be evaluated from line-of-sight velocities, $v_{los}$, and positions, $R_i$, of stars. 
The full data set is $D=\{D_B, D_K \}$ and the posterior probability of our complete data set is:
\begin{equation}\label{DLI_BSplines_posterior}
P(\theta | D)\propto P(\theta) \mathcal{L}(D|\theta) 
\end{equation}
 $P(\theta)$ represents  the probability of uniform prior range for each variable, i.e.:
 \begin{equation}
P(\theta)=\prod_{i=1}^{N_{\text{params}}} \frac{1}{\Delta \theta_i}, 
 \end{equation}
when $\theta \in \Delta \theta_i$ and $0$ otherwise. 
 $N_{\text{params}}$ represents the total number of parameters and $\Delta \theta_i$ the range of possible values for parameter $i$.  
$\mathcal{L}(D|\theta)$ is the likelihood model.

Our likelihood model must take into account both the brightness and kinematic data. Since these two datasets are mutually independent it follows that: 
\begin{equation}\label{DLI_BSplines_lkhoodAnal_4}
\mathcal{L}(D|\theta)=\mathcal{L}(D_B|\theta) \mathcal{L}(D_K|\theta)
\equiv \mathcal{L}_B \cdot \mathcal{L}_K
\end{equation}
For $\mathcal{L}_B$ and $\mathcal{L}_K$ we choose standard Gaussian distributions, i.e.:
\begin{align}\label{DLI_BSplines_lkhoodAnal_5}
 \mathcal{L}_B &= \prod_{i=1}^{N_{\text{data}}}
 \frac{1}{\sqrt{2 \pi (\delta J_i)^2} }
\exp \left(-\frac{(J_i - \Sigma(R_i)/\Upsilon)^2}{2 (\delta J_i)^2} \right) \\
\label{DLI_BSplines_lkhoodAnal_6}
\mathcal{L}_K &=\prod_{i=1}^{N_{\text{data}}}
\frac{1}{\sqrt{2 \pi (\delta d_i)^2} }
\exp \left( -\frac{(d_i - \sigma_{los}^2(R_i))^2}{2 (\delta d_i)^2} \right).
\end{align}
$J_i$ is our brightness data values, $\delta J_i$ the error in each value. $d_i$ is the line-of-sight $\sigma_{los}^2(R_i)$ data value at  position $R_i$, $\delta d_i$ the corresponding error.   In the example presented in Section \ref{DLI_BSplines_Examples}, the brightness 
$J_i$ and line-of-sight velocity dispersion observables $d_i$ are evaluated on the same positions, $R_i$, however 
this need not be the case and this does not affect the efficiency of the method.

In order to estimate the highest likelihood values of the parameters $\theta$ we employed a Markov Chain Monte Carlo (MCMC) algorithm, namely a stretch move as described in \cite{GoodmanWeare}. This method has the advantage of exploring the parameter space efficiently, and the
fitted parameters generally do not get stuck around local maxima of the likelihood function. This is an important feature since if there is a degeneracy in pair of $(\rho,\beta)$ values, then we must recover multimodal distributions for the parameters $a_i$. 
Our MCMC walks were run for sufficient autocorrelation time, so as to ensure that the distributions of parameters were stabilized around certain values. 

Two important remarks need to be made here: due to the complexity of the problem, if we increase the number of $a_i$ coefficients to more than 15, the autocorrelation time\footnote{Number of points in the MCMC walks that are required for the distributions of parameter values to be stabilized.} becomes very large. There were cases in our initial trial runs where we needed to run our MCMC for up to $10^7$ points, because the chains were converging very slowly. The behaviour of the chains is different than in standard parameter estimates of functions. Specifically the values tend to concentrate at some region quite fast, and then this whole region oscillates slowly until it is eventually stabilized. In general, for our models, we run our MCMC for approximately $2-4 \times 10^6$ points, and this was sufficient. However we did not need to use more than $7-12$ unknown $a_i$ coefficients.

\subsection{Bayesian Model Selection}

\begin{table}
\caption{Jeffreys table}
\centering
\begin{tabular}{@{}  c | c | c }
\hline
$\ln \left( \frac{p(M_1)}{p(M_2)} \right) $ & $\frac{p(M_1)}{p(M_2)}$ & Strength of evidence \\
\hline
\hline
$<0$ & $<1$  & negative (supports M2)\\
\hline
0 to $1.16$ & 1 to 3.2 & barely worth mentioning \\
\hline
$1.16$ to $2.3$ & 3.2 to 10 & positive \\
\hline
$2.3$ to $ 4.6$ & 10 to 100 & strong \\
\hline
$>4.6$ & $>100$ & very strong - decisive\\
\hline
\end{tabular}
\label{DLI_BSplines_Jeffrey}
\medskip

 Jeffreys table is a quantitative table for the comparison of two competing models $M_1$ and $M_2$.
\end{table}

In our present description we  use Bayesian model selection \citep{gelman2003bayesian,2010blda.book.....G} and Nested Sampling (\citealt{Skilling}, hereafter JS04), a method for estimating the evidence for a given likelihood model. For completeness, we give a short introduction to these methods.

Let $M_i$ represent each of the models used in our analysis (e.g. $M_1\equiv$ King mass density, with specific number $n$ of coefficients $a_i$, order of B-spline $k$ and knot distribution). Furthermore let $I=M_1+\cdots + M_n$ represent our hypothesis, that at least one of the models is correct. Summation indicates logical ``or''. Let   $\theta$ represent the total number of parameters for each model and $D$ our data set. According to Bayes theorem the probability  of the model parameters $\theta$ given the data set of values is:
\begin{equation}
p(\theta | D,M_i,I)= \frac{p(\theta|M_i,I) \mathcal{L}(D| \theta, M_i,I)}{p(D|M_i,I)}.
\end{equation}
$p(\theta|M_i,I)$ is the prior information on the parameters,  $\mathcal{L}(D| \theta, M_i,I)$ is the likelihood as defined in Equation  \ref{DLI_BSplines_lkhoodAnal_4}  and $p(D| M_i,I)$ is the normalization constant for the model $M_i$ under consideration. 
This constant plays an important role for model selection. 
Marginalizing over all parameters, for the set of competing hypothesis, the probability of a model given  the data is:
\begin{equation}
p(M_i|D,I) = \frac{p(M_i|I) p(D|M_i,I)}{p(D|I)}
\end{equation}
Our level of ignorance of model choice suggests that $p(M_i|I)=p(M_j|I)$ for any $i,j$ combination (all models are equiprobable). Hence the relative ratio of probabilities of two models is: 
\begin{equation}
\frac{p(M_i|D,I)}{p(M_j|D,I)} = \frac{p(M_i|I) p(D|M_i,I)}{p(M_j|I) p(D|M_j,I)}=
\frac{ p(D|M_i,I)}{p(D|M_j,I)}=O_{ij}
\end{equation}
$O_{ij}$ is defined as the odds ratio, and it quantifies the comparison of two competing models for the description of observables.  
$p(D|I)$ is the normalization constant that does not participate in our calculations each time we compute the relative ratio of two models.  
A  measure for model selection is given by Jeffreys table (Table \ref{DLI_BSplines_Jeffrey}). It quantifies the relative ratio of probabilities $p(M_1)/p(M_2)$ of two competing models. 
See 
\cite{Jeffreys61} and 
\cite{gelman2003bayesian} for further details.

Nested Sampling, introduced by JS04,  is an algorithm for the estimation of the normalization parameter $p(D|M_i,I)$. 
Following his terminology, the evidence $Z_i$ of model $M_i$ is given by:
\begin{equation} \label{Skilling_evidence}
Z_i= p(D|M_i,I) = \int p(\theta|M_i,I) \mathcal{L}(D| \theta, M_i,I) d\theta 
\end{equation}
and corresponds to the normalization constant $p(D|M_i,I)$. 
Making use of the prior mass $dX=p(\theta|M_i,I) d\theta$, an effective parameter transformation from $\dim (\theta) = n$ to $\dim(X) =1$, the above integral is simplified: 
\begin{equation}
Z_i=\int_{0}^1 \tilde{\mathcal{L}}(D|X) dX
\end{equation}
In order to estimate this quantity and perform model selection we use MultiNest \citep{2008MNRAS.384..449F,2009MNRAS.398.1601F}.  This algorithm is designed for effective calculation of Bayesian evidence based on Skilling's algorithm. It gives consistent results even in the case of multimodal likelihood functions.

\section{Example: Isotropic system with King mass density} \label{DLI_BSplines_Examples}

In this section we are going to reconstruct the kinematic profile of an isotropic King model ($\beta(r)=0$). For the notation of the total number $n$ of unknown coefficients $a_i$ we use the following scheme: based on the restriction that all the quantities that describe the cluster must be zero at the tidal radius, the final coefficient will be $a_{n}=0$. This extra coefficient does not go into the likelihood analysis, hence we break the total number of coefficients to the sum of unknowns plus one which represents this last coefficient. We use this notation in the figure captions and in Table \ref{DLI_BSplines_Example1_evidence} where we list the Bayesian evidence.

\begin{table}
\caption{Bayesian Evidence for  isotropic  system with King mass model}
\label{DLI_BSplines_Example1_evidence}
\centering
\begin{tabular}{@{}  c | c | c }
 \hline 
 Order & Number of coefficients $n$  &$\ln Z \pm \delta (\ln Z)$ 
 \\ \hline \hline
\multirow{2}{*}{$k=4$ } & $n=5+1$     & $ -50.15 \pm 0.19$  \\[6pt]  
& $n=6+1$  &   $   -41.19\pm 0.20$  \\[6pt] \hline 
 $k=5$  &  $n=5+1$   & $ -41.44 \pm 0.20 $  \\[6pt] 
\hline
  $k=5$  &$n=6+1$   &  $ -45.06 \pm  0.20 $ 
 \\ \hline 
\end{tabular}
\medskip

From left to right: first column is the order $k$ of the B-spline representation of $\psi\equiv \sigma_{rr}^2(r)$. Second column is the number $n$ of coefficients $a_i$.  The third  column is the value of Bayesian evidence as estimated from MultiNest. The highest value of $\ln Z$ corresponds to the most probable model.  
\end{table}

For each evaluation of our likelihood based on our parameters $\theta = (w_0,\rho_0,r_c,a_1,\ldots,a_{n-1}) $ we need to construct a B-spline representation of the radial velocity dispersion $\psi \equiv \sigma_{rr}^2=\sum_i a_i B_{i,k}(r)$. We keep the order $k$ of the B-spline basis fixed. However we use an adaptive knot distribution. For each set of proposed  parameters that define the dynamical model $(w_0,\rho_0,r_c)$, there exists a unique tidal radius $r_t$. This is defined from the solution of the Poisson equation and depends only on the brightness profile. 
We define the knot sequence $\xi_i$ from a Gaussian distribution of knots, around the mean $\mu=r_c$ for the choice of $\gamma=3$ in the interval $r\in[0,r_t]$ 
(section \ref{DLI_BSplines_knot_section}). The choice for the value of $\gamma$ was taken after many tests on synthetic data, by comparison of the Bayesian evidence for each fit. We use a mild concentration of knot points around $r_c$
since we know from theory that this is a point of interest in the sense of increased curvature of the corresponding functions. We need in general more knots  around regions of increased curvature. We emphasise, however, that we can recover the correct kinematic profile also with a uniform knot distribution.

\begin{figure}
\centering
\includegraphics[width=\columnwidth]{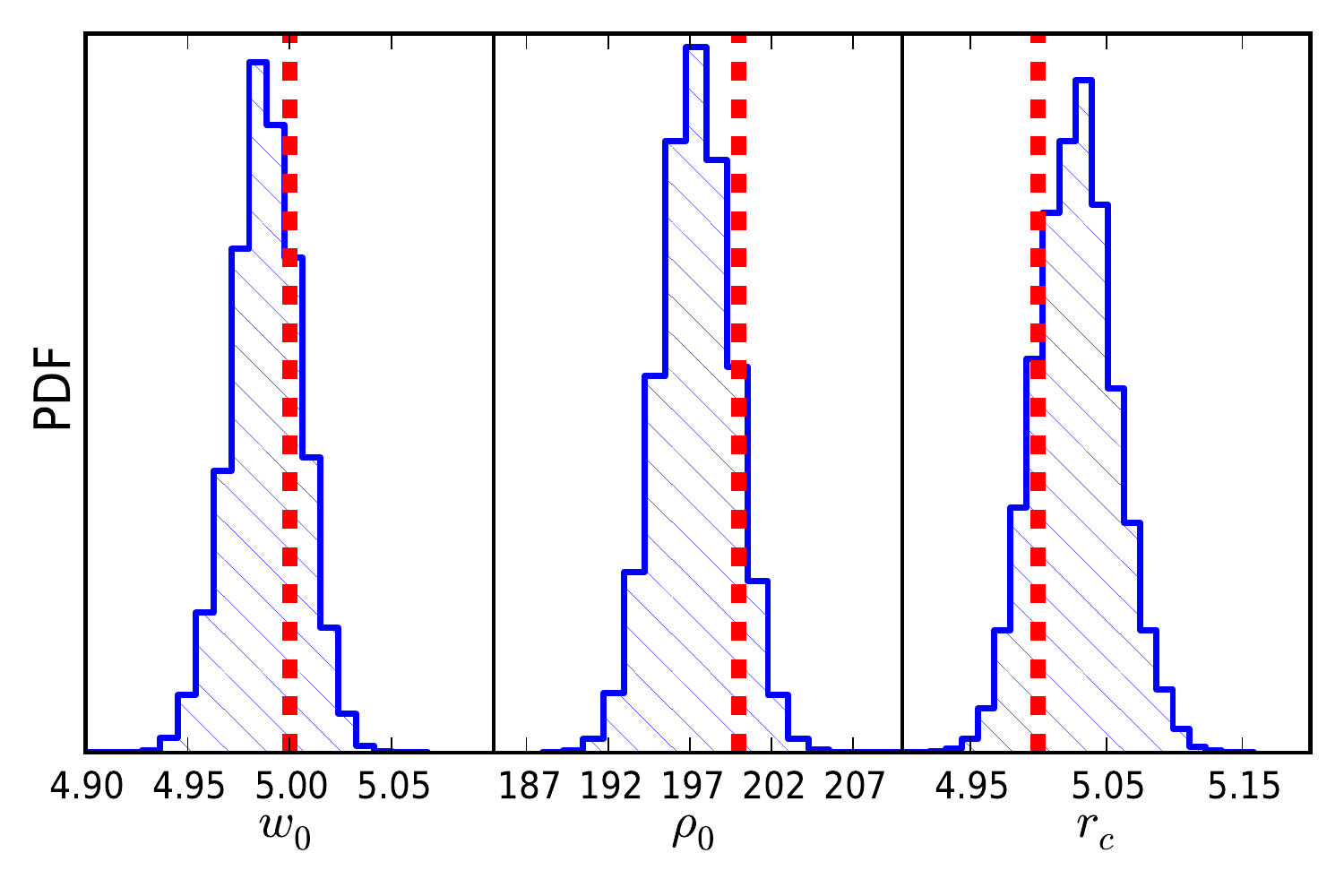}
 \caption{Marginalized distributions of  mass model parameters for the case of   an
 isotropic system with  King mass density and  14 data points, as estimated from our algorithm. In this case we kept fixed the mass-to-light ratio. The fit corresponds to $n=5+1$ coefficients $a_i$.  The red dashed lines correspond to the reference values $\{ w_0^{ref},\rho_0^{ref},r_c^{ref}\}=\{5,200,5\}$ from which synthetic data were created. The reference values of the mass model are well within the boundaries of the estimated values.}
\label{DLI_BSplines_Fig_mass_params}
\end{figure}

\begin{figure*}
\centering
\includegraphics[width=\textwidth]{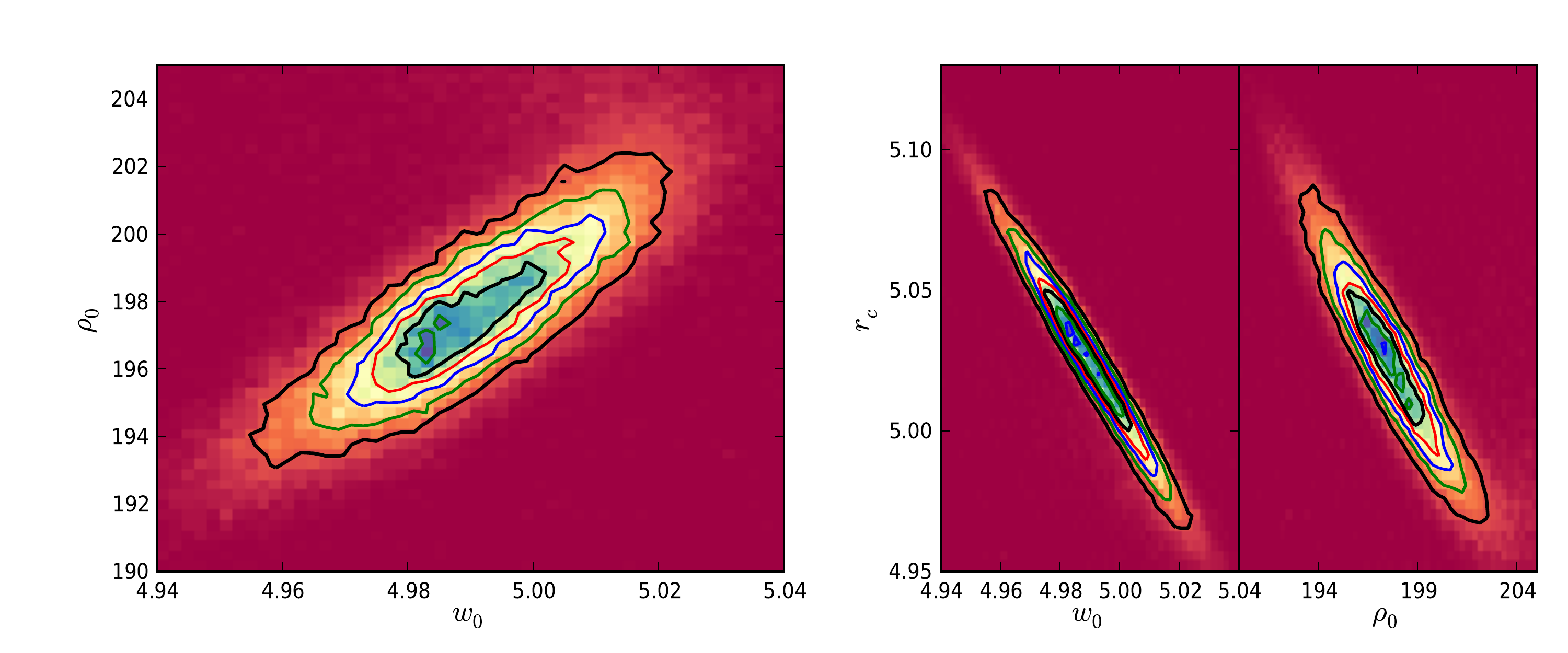}
 \caption{Density plots of the MCMC walks for the defining parameters $(w_0,\rho_0,r_c)$  of the King mass model.}
\label{DLI_BSplines_Fig_mass_density_plots}
\end{figure*}

We use only 14 data points; this small number is used in 
order to demonstrate the power of the method for realistic applications. 
Our data consists of synthetic  brightness values  $J(R)=\Sigma(R)/\Upsilon$ and synthetic line-of-sight velocity dispersion values $\sigma_{los}^2$. For simplicity  we assume a fixed mass-to-light ratio $\Upsilon = 1$ and the following set of values for the defining parameter of the King reference profile: $\{ w_0=5, \rho_0 = 200, r_c = 5\}$.  
Each of the synthetic profile values is constructed by adding a random error to the reference profile, either to the brightness $J$ or to the line-of-sight velocity dispersion $\sigma_{los}^2$. For this  example  the error is a random $10\%$ on the  actual value of the reference profile.  The set of random values is created by the following scheme: 
\begin{align}
J_i &= J(R_i) + \delta J_i, & \delta J_i &= 0.1 J(R_i) g_1\\
d_i &= \sigma_{los}^2(R_i) + \delta d_i, &   \delta d_i &= 0.1 \sigma_{los}^2(R_i) g_2
\end{align}
where $g_1, g_2$ are two distinct Gaussian random numbers of mean zero and dispersion equal to one. $\delta J_i$ is the random error on the brightness value $J_i$, and $\delta d_i$ is the random error of the line-of-sight velocity dispersion $\sigma_{los}^2(R_i)$ at position $R_i$.
 
The first thing we need to establish is if the mass content of the system is recovered correctly. In order to have the most accurate results, we run MultiNest for the evaluation of Bayesian evidence for a set of values for the order $k$ of the B-spline basis and the number of coefficients. The values of all these parameters can be seen in Table \ref{DLI_BSplines_Example1_evidence}. 
Optimum choices result for order $k=4$ and $n=6+1$ coefficients  and order $k=5$ and $n=5+1$ coefficients.  Since these values are the same within error estimates, we choose for our fits the  model with  $k=5$ and $n=5+1$.  For this choice of parameters, we plot the highest likelihood fitting models.

In Fig. \ref{DLI_BSplines_Fig_mass_params} we plot the histograms of the mass model defining parameters $\{w_0,\rho_0,rc\}$. Despite the fact that we allowed complete freedom\footnote{The prior range for all coefficients $a_i$ that define $\sigma_{rr}^2$ was in the range $[0,50]$} in the kinematic profile ($\sigma_{rr}^2$) through the B-spline representation, the recovered parameters of the mass model are distributed around the reference values $w_0^{\text{ref}}=5$,   
 $\rho_0^{\text{ref}}=200$ and $r_c^{\text{ref}}=5$. This is the most important result: the mass content is completely reconstructed. We note that we could have left the mass-to-light ratio as a free parameter and this would still be recovered (Paper II). 
 In Fig.  \ref{DLI_BSplines_Fig_mass_density_plots} we plot the density plots of the MCMC walks for these parameters. 

\begin{figure*}
\centering
\includegraphics[width=\textwidth]{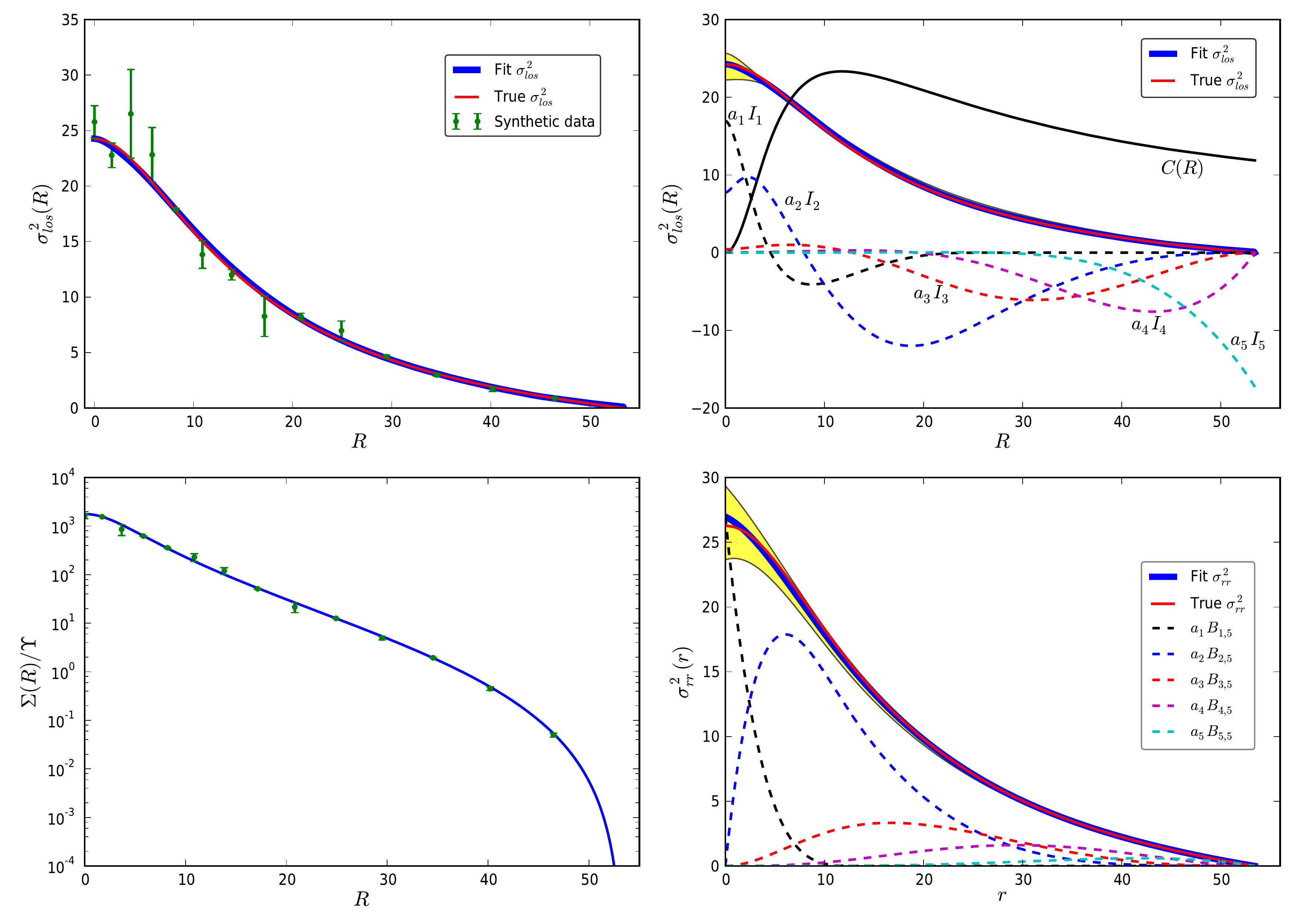}
 \caption{Top left panel: Synthetic line-of-sight velocity dispersion data,  $\sigma_{los}^2$  reference value (red line) and fit (blue line) from the solution of the SSJE using $n=5+1$ coefficients $a_i$ for the B-spline representation of $\sigma_{rr}^2$.  The order of the B-spline basis is $k=5$ and the fit is for  14  synthetic data points. 
Bottom left panel: Fit of the brightness profile of the mass model. Top right panel: 
$\sigma_{los}^2$  reference value (red line) and fit (blue line). We also plot the functions $C(R)$ (Equation \ref{DLI_BSplines_C}) and the weighted $a_i I_i(R)$ (Equation \ref{DLI_BSplines_I}) that define the value of $\sigma_{los}^2$ (Equation \ref{DLI_BSplines_fundamental}).  
 Bottom right panel: the theoretical $\sigma_{rr}^2$ from the King  profile and from the B-spline representation, as estimated by the MCMC procedure. We also plot the weighted B-spline basis $a_i B_{i,k}(x)$. The yellow shaded region in all panels corresponds to $1\sigma$ uncertainty intervals of the coefficients $a_i$ keeping the defining parameters of the mass model fixed to the highest likelihood values. }
\label{DLI_BSplines_Fig2a}
\end{figure*}

Next we need to see how well the kinematic profile, i.e. $\sigma_{rr}^2$ and $\sigma_{los}^2$, is approximated\footnote{Once $\sigma_{rr}^2$ is known, the tangential $\sigma_{tt}^2$ can be evaluated from the SSJE.}.  
In the top left panel of Fig. \ref{DLI_BSplines_Fig2a}  we plot the line-of-sight velocity dispersion $\sigma_{los}^2$ of the synthetic data, the true curve, and the fitted curve. It can be seen that the fit is excellent. In the bottom left panel we plot the synthetic brightness data and the highest likelihood fitting profile. 
The simultaneous fit to brightness and kinematic data is excellent;  this  is important for cases with real data sets where we need to consider the brightness fit as well.

The top right panel deserves special attention: we plot again the line-of-sight velocity dispersion and the $1\sigma$ uncertainty interval (yellow shaded region) that corresponds to the uncertainty in the estimates of $a_i$, and does not account for the variance of the mass model parameters. The region close to the origin ($r=0$) demonstrates greater variance. This is  to be expected, since there the data points end, and the coefficient 
$a_1=\sigma_{rr}^2(0)$ exhibits greater variance. This is a result of the local modification scheme of the B-spline basis functions. Recall that at each point $x\in [\xi_i,\xi_{i+1})$ only a finite number of B-spline bases $B_{i,k}(x)$ are non zero. For the case of $a_1$, since it is at the beginning of the knot distribution, the only non zero B-spline basis is $B_{1,5}$. For all other intermediate points, more than one coefficient $a_i$ contributes to the curve estimate, thus the variance is smaller.   As we move towards the tidal radius of the system $r_t$, the variance of the fitted values decreases. This results from the smoothness of B-spline functions as well as the boundary condition that the curve $\mathbf{P}(r)=\left(r,\sigma_{rr}^2\right)$ must pass through the point $(r_t,0)$.

In the same panel we plot the weighted functions $a_i I_i(R)$ and $C(R)$.  $C(R)$ depends only on the mass density of the system (Equation \ref{DLI_BSplines_C}) and not in any way on the kinematic profile. This function rises up positive and then falls slowly asymptotically. On the other hand, the majority of the weighted functions\footnote{Recall that  $a_i \geq 0$ from the restriction that $\sigma_{rr}^2 \geq 0$, since  $\sigma_{rr}^2=\sum_i a_i B_{i,k}(r) $ and $B_{i,k}(r) \geq 0$.
Actually this is a stronger constraint than necessary: it is possible to have some $a_i<0$ and still the radial velocity profile be positive. 
} acquire negative values for $R> 10$. Then the line-of-sight profile is constructed because the weighted functions $a_iI_i$ are subtracted from $C(R)$. That is, the coefficients $a_i$ are regulated by the fact that they must reduce the value of the sum 
$\sum_i a_i I_i(R) + C(R)$ in a smooth way until it drops to zero at $r_t$. This means that the mass profile defines the behaviour of the majority of coefficients $a_i$ away from the origin.  Again, this results from  the smoothness of B-spline functions and the requirement that they must pass through the point $(r_t,0)$. 
This result is general, and does not depend on the specific B-spline representation of $\sigma_{rr}^2$. 

In the bottom right panel we plot the highest likelihood $\sigma_{rr}^2$ and the true value. The fit is again excellent. The yellow shaded region corresponds to the  $1\sigma$ uncertainty interval of the $a_i$ coefficients only. In the same panel we plot the weighted B-spline basis $a_i B_{i,5}(x)$. Their linear combination constructs  $\sigma_{rr}^2$. Observe again that due to the smooth behaviour of the B-spline functions, and the requirement that all quantities drop to zero at $r=r_t$, the variance of the fitted values goes to zero as we move away from the system center.

\begin{figure*}
\centering
\includegraphics[width=\textwidth]{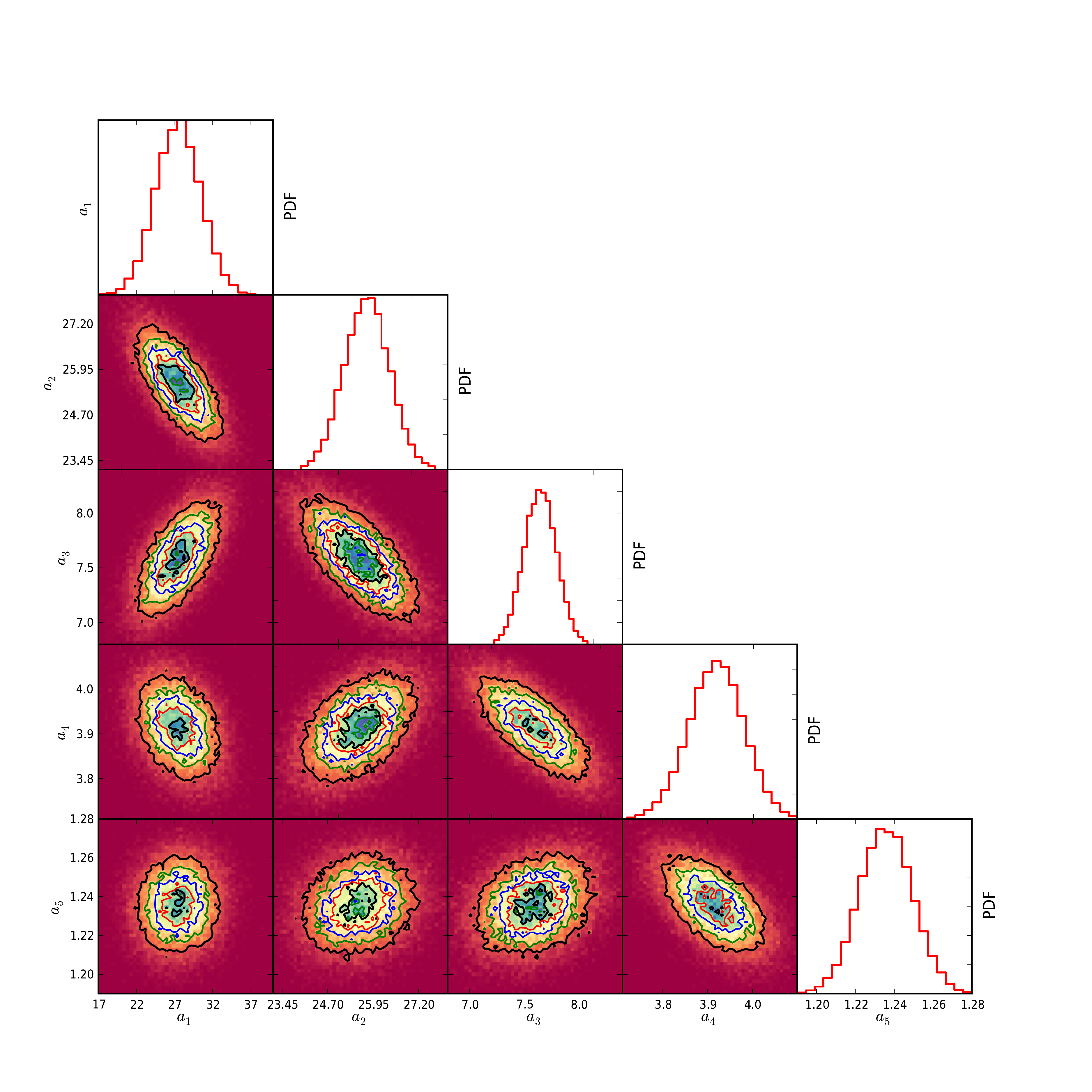}
 \caption{Density plots of the MCMC walks for the parameters $a_i$ in the B-splines representation of $\sigma_{rr}^2=\sum_i a_i B_{i,k}(x)$ for the  isotropic  system with a King  mass model. For this fit we used 14  synthetic data points; the order of the B-spline basis is $k=5$.  }
\label{DLI_BSplines_Fig3}
\end{figure*}

In Fig. \ref{DLI_BSplines_Fig3} we plot the MCMC density plots for the various parameters $a_i$ of the B-spline representation of $\sigma_{rr}^2$. As the index $i$ increases the coefficients $a_i$ correspond to control points further away from the cluster center. Their variance, as expected and as it is evident from the density plots, is reduced away from the cluster center.

\section{Discussion}
\label{DLI_BSplines_Discussion}

Having established the basic mathematical framework of our method, there remain some more issues  to be addressed. 
An important question is how many coefficients $a_i$ can we use? These cannot be arbitrary in number, as our choice is limited by the number of available data. This results from the local modification scheme of B-spline functions: if we use a very large number of coefficients, we end up with regions of the domain of definition of the B-spline function between data points, that are regulated by some $a_i$ coefficients that do not participate in the likelihood function. Bayesian model inference can solve this problem,  since it heavily penalizes models with increased complexity (greater number of coefficients). 
In fact, this is so important that we cannot apply our method without using model selection, Bayesian or frequentist (e.g. Generalised Cross Validation (GCV)).
It may also be the case that our data are noisy and result in kinematic profiles ($\sigma_{rr}^2$ and  $\sigma_{tt}^2$) that have unphysical variations. This is related to the problem of underfitting, overfitting and optimum smoothing  that we further develop in Paper II. There we will address the issue of optimum smoothing, by defining a curvature penalty on the B-spline representation using information of smoothness from ideal theoretical models.

Another important issue is whether the reconstructed kinematic profile is physically acceptable. 
When we observe a real stellar system, then this possesses a kinematic profile that is a physical realisation, i.e. we cannot question whether $\sigma_{los}^2$ is correct, since this is dictated from nature (although we may question the reliability of the observations!). The question is if the decomposition to $\sigma_{rr}^2$ and $\sigma_{tt}^2$ is physically acceptable for a given mass density $\rho(r)$. For a general single stellar model with a constant mass-to-light ratio, to the point where our assumption of mass density is a good approximation, then it must be. We expect that if we have a good approximation to the real brightness distribution then the kinematic profile must be also in good proximity with reality. 
If any of our assumptions is significantly flawed, then we will observe some unrealistic behaviour in $\sigma_{tt}^2$ through the use of the SSJE.

Is  the kinematic profile that reproduces the line-of-sight velocity dispersion, i.e. $\sigma_{rr}^2$ and $\sigma_{tt}^2$,  unique?
Let us assume that we know a complete theoretical functional form of $\sigma_{los}^2$ and the mass density, $\rho(r)$, of a self gravitating  system.
In principle it is possible to find $n$   positions  $R_j$ and form a linear system from Equation \ref{DLI_BSplines_fundamental} for the unknowns $a_i$ ($n$ in total). Care must be taken, since $R_j$ positions must span all of the distance $[0,r_t]$ for the system to have a solution. A natural choice is to use the collocation points (Greville abscissae, see section \ref{DLI_paperI_BSpline_functions}) for the $R_j$. The matrix $I_i(R_j)$ that is produced in this way, is a band matrix due to the local modification scheme of B-spline bases. 
 Then, if this matrix $I_i(R_j)$ is invertible,  solving for $a_i$ results in a unique solution to the system. It is important to note here that, a given knot distribution will result a unique solution, however this solution may be far from optimum. It will trace the correct $\sigma_{los}^2$ and $\sigma_{rr}^2$ profiles, but it may have non physical variational behaviour. Using an optimization algorithm (e.g. Genetic Algorithm)  for the number of coefficients, $a_i$,  positions of knots, $\xi_i$, and order, $k$, of the B-spline basis, it is in principle possible to find a solution for the unknowns $a_i$ and knot distribution that will give in a desirable accuracy the kinematic profile $\sigma_{los}^2$, and in consequence, $\sigma_{rr}^2$. Within this desired accuracy, the kinematic profile will be unique.  
That is, for a given line-of-sight velocity dispersion $\sigma_{los}^2(R)$ and a given potential-mass density pair  $(\Phi,\rho)$, if and only if  $I_i(R_j)$ is invertible, there exists a unique decomposition to $\sigma_{rr}^2$ and $\sigma_{tt}^2$. 
 Therefore, there exists a unique anisotropy profile.   
The fact that the B-spline functions are an approximation to the true function $\sigma_{rr}^2$ should not worry us. B-spline bases form a set of ``as complete as possible'' bases for the representation of a function. Actually, in the limit where $n\to \infty$ a B-spline representation of a function is its Taylor expansion around some position $x$ \citep{de1978practical}. 
For the case where we have discrete data we cannot speak of a unique profile. Rather  we have a unique family of profiles, within the statistical uncertainty of the model parameters as this is estimated from the MCMC.  If the system is degenerate, using a Markov chain exploration of parameter space will in general result in multiple peaks in the marginalised distributions of the $a_i$ coefficients and the defining parameters of the mass models.

\section{Conclusions}
\label{DLI_BSplines_Conclusions}

In this paper we describe the basic mathematical framework for the removal of the mass-anisotropy degeneracy of the SSJE. This is achieved to the level that for an assumed functional form of the  pair  $(\Phi(r), \rho(r))$ of potential and mass density, and a given  data set of brightness $J$ and line-of-sight velocity dispersion observables $\sigma_{los}^2$,  we reconstruct a unique kinematic profile $(\sigma_{rr}^2,\sigma_{tt}^2)$, within the statistical uncertainties. The uniqueness of this follows from an exploration of parameter space through an MCMC scheme, i.e. we do not present a formal mathematical proof of why this is the case. 

Our algorithm  combines smoothing B-splines with dynamical equations of physical systems and reconstructs accurately the kinematic profile and the mass content of a stellar system. This is for a constant or variable mass-to-light ratio $\Upsilon$. In the current contribution we present a simple example of an isotropic King profile with fixed mass-to-light ratio.  We explore models with variable $\Upsilon$ in Paper II.

Finally, we note  that the idea of combining smoothing splines with equations from dynamics and allowing statistical inference to give the correct shape of unknown functions is quite general. It can be applied to any system in which there are physical quantities of unknown theoretical functional form where we wish to determine their approximate shape. This avoids bias in model parameter estimates and allows for a better understanding of physical models.\\

\section*{Acknowledgments}
F. I. Diakogiannis acknowledges the University of Sydney International Scholarship
for the support of his candidature. 
G. F. Lewis acknowledges support from ARC Discovery Project (DP110100678) and Future Fellowship (FT100100268). 
The authors would like to thank Nick Bate  as well as the anonymous referee for useful comments and suggestions on the manuscript.   

%
\bibliographystyle{mn2e} 
\bibliography{DLI_BSplines_I.bib} 

\begin{thebibliography}{}

\bibitem[\protect\citeauthoryear{{Binney} \& {Mamon}}{{Binney} \&
  {Mamon}}{1982}]{1982MNRAS.200..361B}
{Binney} J.,  {Mamon} G.~A.,  1982, \mnras, 200, 361

\bibitem[\protect\citeauthoryear{{Binney} \& {Tremaine}}{{Binney} \&
  {Tremaine}}{2008}]{2008gady.book.....B}
{Binney} J.,  {Tremaine} S.,  2008, {Galactic Dynamics: Second Edition}.
Princeton University Press

\bibitem[\protect\citeauthoryear{De~Boor}{De~Boor}{1978}]{de1978practical}
De~Boor C.,  1978, A Practical Guide to Splines.
No. v. 27 in Applied Mathematical Sciences, Springer-Verlag

\bibitem[\protect\citeauthoryear{{Dejonghe} \& {Merritt}}{{Dejonghe} \&
  {Merritt}}{1992}]{1992ApJ...391..531D}
{Dejonghe} H.,  {Merritt} D.,  1992, \apj, 391, 531

\bibitem[\protect\citeauthoryear{Farin}{Farin}{2002}]{Farin:2001:CSC:501891}
Farin G.,  2002, Curves and surfaces for CAGD: a practical guide, 5th edn.
Morgan Kaufmann Publishers Inc., San Francisco, CA, USA

\bibitem[\protect\citeauthoryear{{Feroz} \& {Hobson}}{{Feroz} \&
  {Hobson}}{2008}]{2008MNRAS.384..449F}
{Feroz} F.,  {Hobson} M.~P.,  2008, \mnras, 384, 449

\bibitem[\protect\citeauthoryear{{Feroz}, {Hobson} \& {Bridges}}{{Feroz}
  et~al.}{2009}]{2009MNRAS.398.1601F}
{Feroz} F.,  {Hobson} M.~P.,    {Bridges} M.,  2009, \mnras, 398, 1601

\bibitem[\protect\citeauthoryear{Gelman, Carlin, Stern \& Rubin}{Gelman
  et~al.}{2003}]{gelman2003bayesian}
Gelman A.,  Carlin J.,  Stern H.,    Rubin D.,  2003, Bayesian Data Analysis,
  Second Edition.
Chapman \& Hall/CRC Texts in Statistical Science, Taylor \& Francis

\bibitem[\protect\citeauthoryear{{Gerhard}}{{Gerhard}}{1993}]{1993MNRAS.265..213G}
{Gerhard} O.~E.,  1993, \mnras, 265, 213

\bibitem[\protect\citeauthoryear{{Goodman} \& {Weare}}{{Goodman} \&
  {Weare}}{2010}]{GoodmanWeare}
{Goodman} G.~G.,  {Weare} J.~J.,  2010, CAMCoS, 5, 65

\bibitem[\protect\citeauthoryear{{Gregory}}{{Gregory}}{2010}]{2010blda.book.....G}
{Gregory} P.,  2010, {Bayesian Logical Data Analysis for the Physical Sciences}

\bibitem[\protect\citeauthoryear{Hastie, Tibshirani \& Friedman}{Hastie
  et~al.}{2001}]{hastie01statisticallearning}
Hastie T.,  Tibshirani R.,    Friedman J.,  2001, The Elements of Statistical
  Learning.
Springer Series in Statistics, Springer New York Inc., New York, NY, USA

\bibitem[\protect\citeauthoryear{{Ibata}, {Nipoti}, {Sollima}, {Bellazzini},
  {Chapman} \& {Dalessandro}}{{Ibata} et~al.}{2013}]{2013MNRAS.428.3648I}
{Ibata} R.,  {Nipoti} C.,  {Sollima} A.,  {Bellazzini} M.,  {Chapman} S.~C.,
  {Dalessandro} E.,  2013, \mnras, 428, 3648

\bibitem[\protect\citeauthoryear{Jeffreys}{Jeffreys}{1961}]{Jeffreys61}
Jeffreys H.,  1961, Theory of Probability, third edn.
Oxford, Oxford, England

\bibitem[\protect\citeauthoryear{{King}}{{King}}{1966}]{1966AJ.....71...64K}
{King} I.~R.,  1966, \aj, 71, 64

\bibitem[\protect\citeauthoryear{{{\L}okas}}{{{\L}okas}}{2002}]{2002MNRAS.333..697L}
{{\L}okas} E.~L.,  2002, \mnras, 333, 697

\bibitem[\protect\citeauthoryear{{Merrifield} \& {Kent}}{{Merrifield} \&
  {Kent}}{1990}]{1990AJ.....99.1548M}
{Merrifield} M.~R.,  {Kent} S.~M.,  1990, \aj, 99, 1548

\bibitem[\protect\citeauthoryear{{Merritt}}{{Merritt}}{1987}]{1987ApJ...313..121M}
{Merritt} D.,  1987, \apj, 313, 121

\bibitem[\protect\citeauthoryear{{Navarro}, {Frenk} \& {White}}{{Navarro}
  et~al.}{1996}]{1996ApJ...462..563N}
{Navarro} J.~F.,  {Frenk} C.~S.,    {White} S.~D.~M.,  1996, \apj, 462, 563

\bibitem[\protect\citeauthoryear{Rogers}{Rogers}{2001}]{Rogers:2001:INH:347021}
Rogers D.~F.,  2001, An introduction to NURBS: with historical perspective.
Morgan Kaufmann Publishers Inc., San Francisco, CA, USA

\bibitem[\protect\citeauthoryear{Sivia \& Skilling}{Sivia \&
  Skilling}{2006}]{sivia2006data}
Sivia D.,  Skilling J.,  2006, Data analysis: a Bayesian tutorial.
Oxford science publications, Oxford University Press

\bibitem[\protect\citeauthoryear{{Skilling}}{{Skilling}}{2004}]{Skilling}
{Skilling} J.~J.,  2004, in Instrumentation and Research Programmes for Small
  Telescopes. p.~395

\bibitem[\protect\citeauthoryear{{Tonry}}{{Tonry}}{1983}]{1983ApJ...266...58T}
{Tonry} J.~L.,  1983, \apj, 266, 58

\bibitem[\protect\citeauthoryear{{van der Marel}}{{van der
  Marel}}{1994}]{1994MNRAS.270..271V}
{van der Marel} R.~P.,  1994, \mnras, 270, 271

\bibitem[\protect\citeauthoryear{{Wolf}}{{Wolf}}{2011}]{2011IAUS..271..110W}
{Wolf} J.,  2011, in {Brummell} N.~H.,  {Brun} A.~S.,  {Miesch} M.~S.,
  {Ponty} Y.,  eds,  IAU Symposium Vol. 271, IAU Symposium. pp 110--118

\end{thebibliography}

\appendix
\section{Our Exceptional Milky Way} \label{DLI_BSplines_MW}

Relative to much of the halo, our Solar System resides close to the centre of our Galaxy. As a result, many
observable stellar velocities are approximately radial velocities $v_r$. We need then to adapt our formalism to the fact that the direct observable quantity is now $\sigma_{rr}^2$. 
Here we simply demonstrate how to modify the SSJE, in order to be able to compare observables with $\sigma_{rr}^2$. Our goal is not to perform a complete dynamical analysis of the Milky Way potential and stellar structure. 

We are going to use only Equation  \ref{DLI_BSplines_Jeans_psi}. 
Multiplying with the integrating factor $r^2 \rho(r)$ yields: 
\begin{align}
-r^2 \rho \frac{d\Phi}{dr} &= \frac{d (r^2 \rho \psi)}{dr} - r \rho \phi. 
\end{align}
That is, 
\begin{equation} \label{DLI_BSplines_A1}
\psi = \frac{1}{r^2 \rho} \int_{0}^r r \rho \phi dr 
- \frac{1}{r^2 \rho}
\int_{0}^r r^2 \rho \frac{d \Phi}{dr} dr. 
\end{equation}
Where we consider that $r^2 \rho \psi \big|_{r=0} = 0 $, since $ \rho(r)\psi(r)$
 takes  finite positive values at the origin. In cases where $\rho(r)$ is singular at $r=0$, multiplying with $r^2$ can in general remove this singularity. 

We expand the tangential velocity dispersion in a finite B-spline basis, according to: 
\begin{equation}\label{DLI_BSplines_A2}
\sigma_{tt}^2=\phi= \sum_i b_i B_{i,k}(r)
\end{equation}
where $b_i$ are constant coefficients to be determined, and define the shape of
 $\sigma_{tt}^2$. 
Then, the radial velocity dispersion $\sigma_{rr}^2=\psi(r)$ is now expressed in terms of the unknown coefficients $b_i$ of $\sigma_{tt}^2 \equiv \phi(r)$ and the mass model. Substituting Equation \ref{DLI_BSplines_A2}   in Equation \ref{DLI_BSplines_A1} and 
collecting terms, in a similar process as in section \ref{DLI_BSplines_Jeans_section}, yields:  
\begin{equation} \label{DLI_BSplines_MW_psi}
\psi = \sum_{i=1}^n b_i \tilde{I}_i (r) + \tilde{C}(r)
\end{equation}
where 
\begin{align}
\tilde{I}_i (r) & = \frac{1}{r^2 \rho} \int_0^r r \rho B_{i,k}(r) dr \\
\tilde{C}(r) &= - \frac{1}{r^2 \rho} \int_0^r r^2 \rho \frac{d \Phi}{dr} dr. 
\end{align}
Comparison of Equation \ref{DLI_BSplines_MW_psi} with observables, can result in marginalized distributions for the defining parameters of the mass model as well as the unknown coefficients $b_i$ that define the shape of $\sigma_{tt}^2 \equiv \phi$. Back substitution   
into Equation \ref{DLI_BSplines_MW_psi} results in the corresponding radial velocity dispersion $\sigma_{rr}^2 \equiv \psi(r)$. 

Note that although in our observables we do not have information on the tangential velocity distribution, this knowledge is acquired through the use of the SSJE. Indeed, knowing $\sigma_{tt}^2$ or $\sigma_{rr}^2$, we can calculate one from the other. Also, due to the higher number of data points compared to the  unknown parameters, Equation \ref{DLI_BSplines_MW_psi} can be used for estimation of the kinematic profile  and  of the mass model.

\bsp

\label{lastpage}

\end{document}